\documentclass[10pt,conference]{IEEEtran}
\IEEEoverridecommandlockouts\IEEEpubid{\makebox[\columnwidth]{This conference paper was accepted by ICC 2025.\hfill} \hspace{\columnsep}\makebox[\columnwidth]{ }}
\usepackage{mathrsfs}
\usepackage{pifont}
\usepackage{bbding}
\usepackage{tipa}
\usepackage{color}
\usepackage{cite}
\usepackage{epsfig}
\usepackage{graphicx}
\usepackage{url}
\usepackage{amsfonts}
\usepackage{multicol}
\usepackage{float}
\usepackage{times}
\usepackage{cite}
\usepackage{psfrag}
\usepackage{subfigure}
\usepackage{stfloats}
\usepackage{amsmath}
\usepackage{enumitem}

\usepackage{footnote}
\usepackage{array}
\usepackage{amsmath,epsfig}
\usepackage{stmaryrd}
\usepackage{amssymb}
\usepackage{epic}
\usepackage{graphicx}
\usepackage{curves}
\usepackage{balance}
\usepackage{algorithm}
\usepackage{algorithmic}
\usepackage{multirow}
\usepackage{amsmath}
\usepackage{xcolor}
\usepackage{graphics}
\usepackage{epsfig}
\usepackage{epstopdf}
\usepackage{enumerate}
\usepackage{color}
\usepackage{siunitx}
\usepackage{tabularx}
\usepackage{booktabs}

\def\BibTeX{{\rm B\kern-.05em{\sc i\kern-.025em b}\kern-.08em
    T\kern-.1667em\lower.7ex\hbox{E}\kern-.125emX}}
\begin{document}


\title{JPPO: Joint Power and Prompt Optimization for Accelerated Large Language Model Services}
\author{Feiran You\textsuperscript{1}, Hongyang Du\textsuperscript{1}, Kaibin Huang\textsuperscript{1},~\emph{Fellow,~IEEE}, and Abbas~Jamalipour\textsuperscript{2},~\emph{Fellow,~IEEE}
\\
\small{\textsuperscript{1}Department of Electrical and Electronic Engineering, University of Hong Kong, Hong Kong SAR, China}\\
\small{\textsuperscript{2}School of Electrical and Computer Engineering, University of Sydney, Sydney, Australia}
\\
{\small{Emails: fryou@eee.hku.hk, duhy@eee.hku.hk, huangkb@eee.hku.hk, a.jamalipour@ieee.org}}
}
\vspace{-2cm}
\maketitle
\vspace{-2cm}

\begin{abstract}
Large Language Models (LLMs) have demonstrated remarkable capabilities in various tasks, leading to their increasing deployment in wireless networks for a wide variety of user services. However, the growing longer prompt setting highlights
the crucial issue of computational resource demands and huge communication load. To address this challenge, we propose Joint Power and Prompt Optimization (JPPO), a framework that combines Small Language Model (SLM)-based prompt compression with wireless power allocation optimization. By deploying SLM at user devices for prompt compression and employing Deep Reinforcement Learning for joint optimization of compression ratio and transmission power, JPPO effectively balances service quality with resource efficiency. Experimental results demonstrate that our framework achieves high service fidelity and low bit error rates while optimizing power usage in wireless LLM services. The system reduces response time by about $17\%$, with the improvement varying based on the length of the original prompt.
\end{abstract}
 
\IEEEpeerreviewmaketitle

\section{Introduction}
\addtolength{\topmargin}{-0.23in}
Large Language Models (LLMs) have revolutionized natural language processing, demonstrating unprecedented capabilities across various tasks~\cite{min2023recent}. As these models increasingly drive intelligent IoT devices and edge computing applications, their deployment over wireless networks to provide services to end-users has become a critical scenario~\cite{10669603}. 
However, this integration faces significant challenges from the computational demands of LLMs and the inherent constraints of wireless communication systems~\cite{jiang2024large}. 

A key issue lies in the growing length of prompts used to elicit advanced reasoning from LLMs~\cite{10705427}, particularly with the advent of techniques like Chain-of-Thought (CoT) prompting~\cite{wei2022chain} and In-Context Learning (ICL)~\cite{li2024long}. 
While these methods substantially enhance LLM performance, they often necessitate prompts that can extend to tens of thousands of tokens~\cite{xue2024repeat}. This trend creates a fundamental trade-off: On the one hand, longer, more sophisticated prompts unlock the full potential of LLMs; on the other, they impose severe burdens on communication bandwidth and computing resources in wireless settings. 
Moreover, the transmission of these extensive prompts consumes significant network resources and introduces considerable latency, potentially undermining the real-time responsiveness crucial for many applications~\cite{xu2024unleashing}. Consequently, the efficient deployment of LLMs in wireless networks demands innovative solutions that can balance prompt inputs and resource efficiency requirements~\cite{friha2024llm}.

\begin{figure}[t]
\centering
\includegraphics[width=0.5\textwidth]{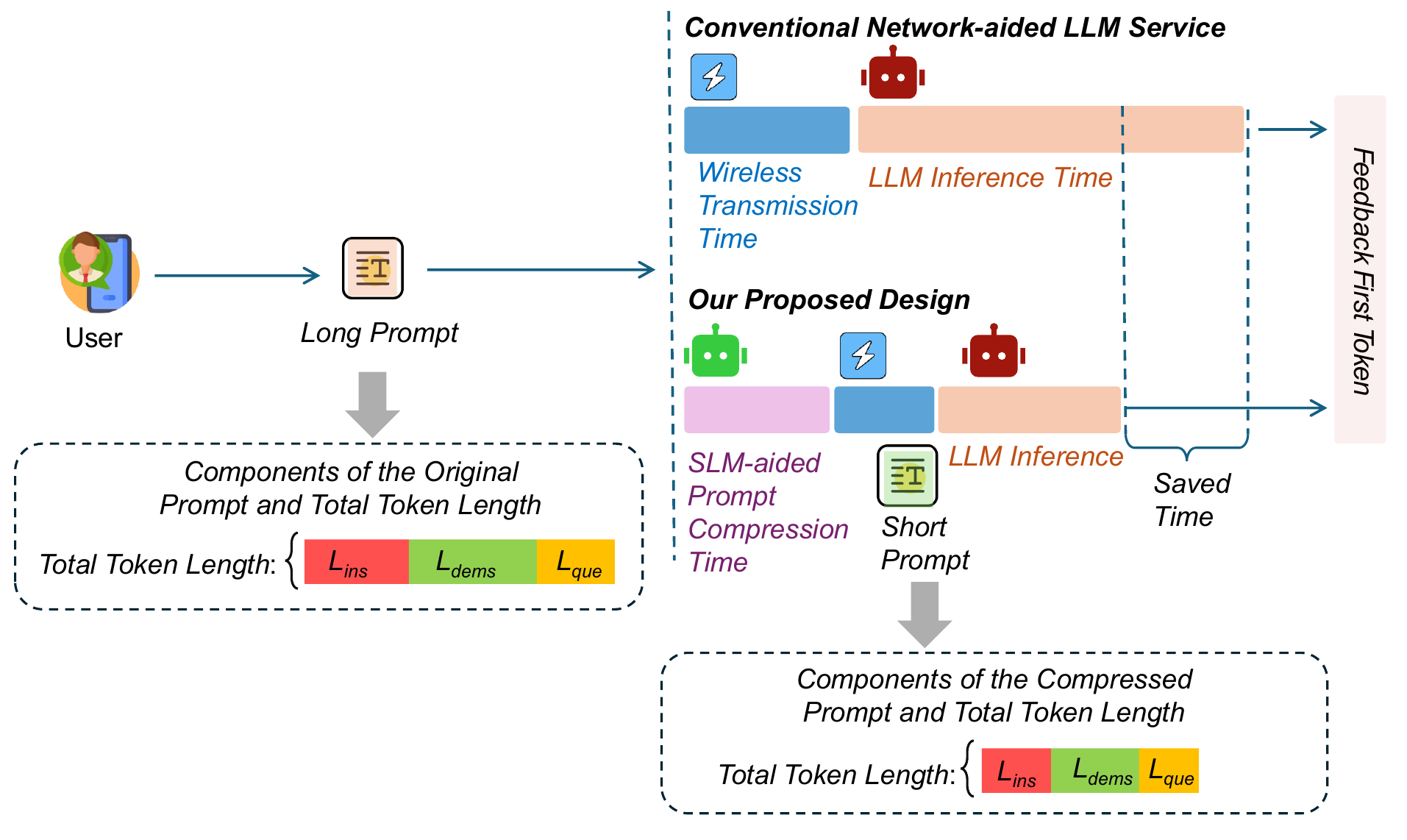}
   \caption{Time consumption comparison of first token generation between conventional network-aided LLM inference service architecture and our design under long user prompts.}
    \label{fig:demo}
\end{figure}
Prior research has tried to address these challenges from different angles. The authors in~\cite{jiang2023llmlingua} introduced LLMLingua, a coarse-to-fine prompt compression method that demonstrates significant potential for compressing LLM prompts while preserving their semantic integrity. LLMLingua utilizes a small language model for compression, which can be aligned with the target LLM through instruction tuning. While this approach shows promise for reducing communication overhead, it does not fully account for the challenges posed by wireless transmission. 
The work in~\cite{liu2024llm} proposes LLM-Slice, a system that creates dedicated network slices for LLM services to improve wireless resource management and reduce response delays. Although LLM-Slice advances wireless network architecture for LLM services, it focuses solely on communication resource allocation without considering prompt optimization for individual user requirements. 
Addressing these issues requires solving two key questions:
\begin{itemize}
    \item {\textbf{Q1)}} How to achieve adjustable prompt compression without significantly degrading LLM inference performance and service quality?
    \item {\textbf{Q2)}} How to jointly design wireless resource allocation, particularly transmission power, to meet the latency and power consumption constraints of network-aided LLM services while maintaining high service quality?
\end{itemize}
For {\textbf{Q1}}, natural language processing methods offer potential solutions for prompt compression. However, traditional NLP compression techniques like text summarization or keyword extraction often fail to capture the complex reasoning patterns and task-specific requirements embedded in LLM prompts, leading to degraded inference performance. Alternative AI-based approaches, such as large autoencoder models or specialized compression networks, require substantial computational resources and introduce additional inference latency at user terminals, making them impractical for resource-constrained wireless scenarios. 
Small language models (SLMs), which can be easily deployed at user terminals, offer a promising solution through their semantic understanding capability to compress LLM prompts while preserving task-critical information. SLM has been adapted as an effective way to employ transformers for edge applications in~\cite{10745806}. Building upon this insight and addressing the {\textbf{Q2}}, we propose Joint Power and Prompt Optimization (JPPO), a framework that combines SLM-based prompt compression with wireless power allocation optimization, as illustrated in Fig. 1. JPPO captures the trade-off between compression ratio and wireless resource consumption, adapting to both channel conditions and prompt content. This integrated approach enables efficient wireless LLM services while maintaining response quality through intelligent prompt optimization. The contributions of this paper are summarized as
\begin{itemize}
    \item We employ a small, aligned language model as a prompt compressor. This approach preserves useful information while significantly reducing prompt size. The SLM efficiently captures essential meaning without requiring training at the transmitter.
\item We formulate an optimization problem for our JPPO framework. This problem balances compression quality, wireless transmission performance, and LLM inference efficiency. 
The objective function is the quality of service (QoS), considering energy limitation and end-to-end latency, with compression ratio and transmission power as key decision variables.
\item To solve the complex JPPO optimization problem, we employ a Deep Reinforcement Learning (DRL) approach. The DRL agent learns to make optimal decisions on compression and transmission strategies, effectively reducing communication overhead and accelerating LLM inference services in latency-constrained and variable channel scenarios.
\end{itemize}

\section{System Model}
In this section, we present the system model of wireless network-aided LLM inference services, the SLM-based prompt compression method, and the wireless transmission model with power consumption and delay constraints.

\begin{figure*}[ht]
\centering\includegraphics[width=0.7\textwidth]{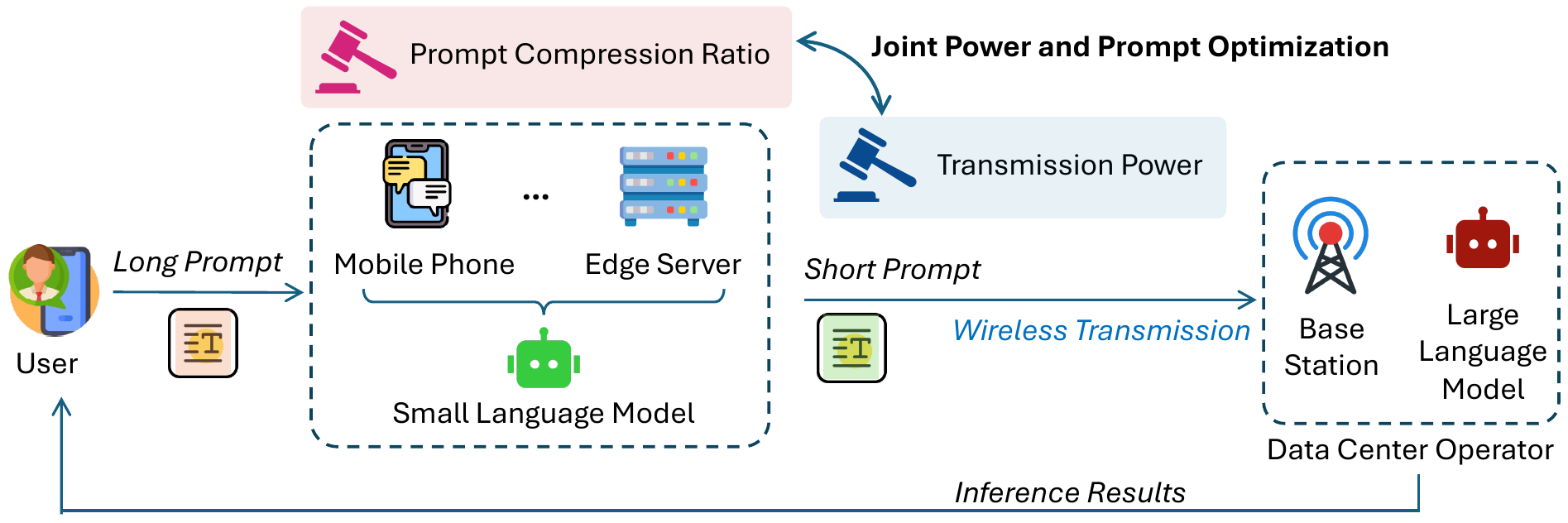}
   \caption{System model of wireless network-aided LLM services and overview of our proposed JPPO, where user-generated long prompts are first compressed through SLM-based edge computing, then transmitted with optimized power allocation via wireless networks to LLM server, and finally inference results are returned to users.}
    \label{fig:system_model}
\end{figure*}

\subsection{Wireless Network-aided LLM Services}

We consider a heterogeneous wireless network where a Data Center Operator (DCO) provides LLM inference services to $N$ users with diverse task requirements, e.g., prompts.
As illustrated in Fig.~2, our proposed framework consists of three key components: an SLM agent deployed at user devices or edge servers for prompt compression, a JPPO scheme for reliable wireless transmission, and a target LLM for inference service. On the user side, the SLM agent leverages its semantic understanding capability to compress prompts while preserving task-critical information. The compressed prompts are then transmitted through wireless channels with jointly optimized power allocation, and finally processed by the target LLM for inference. This framework adaptively adjusts both compression ratios and transmission power based on channel conditions and prompt characteristics to achieve high LLM service quality.

\subsection{Prompt Compression}
\addtolength{\topmargin}{-0.23in}
To efficiently reduce prompt sizes while preserving semantic integrity, we adopt a coarse-to-fine compression approach in SLMs. Our compression method is designed to achieve two primary objectives:
\begin{itemize}
    \item Preserve critical information in the prompt while ensuring the recoverability of the original semantic meaning.
    \item Enable flexible compression ratios that can be dynamically adjusted together with communication resources.
\end{itemize}

Let us formally define the prompt structure and compression process. Consider an original prompt ${\textit{\textbf{x}}}$ that consists of three components:
\begin{equation}
    {\textit{\textbf{x}}} = \left({\textit{\textbf{x}}}_{\text{ins}}, {\textit{\textbf{x}}}_{\text{dems}}, {\textit{\textbf{x}}}_{\text{que}}\right),
\end{equation}
where ${\textit{\textbf{x}}}_{\text{ins}}$ represents the instruction component, ${\textit{\textbf{x}}}_{\text{dems}}$ denotes the demonstrations or examples, and ${\textit{\textbf{x}}}_{\text{que}}$ contains the specific question or task. As depicted in Fig. 1, the token length of each component is denoted by $\mathcal{L}_{\text{ins}}$, $\mathcal{L}_{\text{dems}}$, and $\mathcal{L}_{\text{que}}$ respectively. The total token length $\mathcal{L}_x$ of the original prompt is given by:
\begin{equation}
    \mathcal{L}_x = \mathcal{L}_{\text{ins}} + \mathcal{L}_{\text{dems}} + \mathcal{L}_{\text{que}}.
\end{equation}
Our SLM-based compression mechanism generates a compressed prompt $\hat{\textit{\textbf{x}}}$ with length $\mathcal{L}_{\hat{x}}$. The compression ratio $\kappa$ is defined as:
\begin{equation}
    \kappa = \frac{\mathcal{L}_{\hat{x}}}{\mathcal{L}_x}, \quad \kappa \in [0, 1]
\end{equation}
where $\kappa = 1$ indicates no compression and smaller values of $\kappa$ represent higher compression rates.

We use a comprehensive fidelity metric $\mathbf{f}$ capturing three essential aspects of semantic preservation during information transmission~\cite{10122224} to evaluate the quality of prompt compression. The fidelity metric is composed of three key components:
\begin{itemize}
    \item {\textit{Representation accuracy}} ($\mathbf{f_1}$), which measures how accurately the compressed prompt preserves the semantic meaning of the original prompt and maintains semantic integrity. $\mathbf{f_1}$ is measured by the similarity metric based on the comparison between the representations of the original prompt and the compressed prompt.
    \item {\textit{Transmission completeness}} ($\mathbf{f_2}$), which evaluates the integrity of information preservation during the compression process. $\mathbf{f_2}$ is calculated as the token retained in the compressed prompt compared to the original prompt, considering the transmission channel's Bit Error Rate (BER).
    \item {\textit{Understanding accuracy}} ($\mathbf{f_3}$), which quantifies how well the receiver (target LLM) can interpret the compressed prompt correctly. $\mathbf{f_3}$ is given by the similarity metric to evaluate the proximity of the response to the compressed prompt received and the predefined response to the original prompt.
\end{itemize}

The overall fidelity metric $\mathbf{f}$ can be defined as a weighted sum of these components:
\begin{equation}
    \mathbf{f} = \alpha_1\mathbf{f_1} + \alpha_2\mathbf{f_2} + \alpha_3\mathbf{f_3},
\end{equation}
where $\alpha_1$, $\alpha_2$, and $\alpha_3$ are weight factors determining the relative importance of each fidelity component. These weights can be adjusted based on application requirements and QoS priorities.
\addtolength{\topmargin}{-0.29in}
\subsection{Energy Consumption}
For one user, the total energy consumption $E$ in the one-shot LLM service request process consists of two components: encoding energy $E_e$ and transmission energy $E_t$. This can be expressed as:
\begin{equation}
    E\left(\kappa, P_T\right) = E_e\left(\kappa\right) + E_t\left(\kappa, P_T\right).
\end{equation}
The encoding energy consumption $E_e$, which represents the energy used by the SLM encoder for prompt compression, is calculated as~\cite{faiz2024llmcarbon}
\begin{equation}
    E_e = t_{\text{e}}^{{\rm SLM}} \left( \kappa \right) n_{\text{gpu}}^{{\rm SLM}} P_{\text{gpu}}^{{\rm SLM}} + t_{\text{e}}^{{\rm LLM}} \left( \kappa \right) n_{\text{gpu}}^{{\rm LLM}} P_{\text{gpu}}^{{\rm LLM}},
\end{equation}
where $t_{\text{e}}^{{\rm SLM}}$ represents the GPU execution time in SLM, $n_{\text{gpu}}$ denotes the number of GPUs utilized, and $P_{\text{gpu}}$ is the thermal design power per GPU, and superscript ${\rm LLM}$ denotes the corresponding parameters for the LLM.

The transmission energy consumption $E_t$ is
\begin{equation}
    E_t = t_{\text{t}} \left( \kappa \right) P_{\text{T}} = \frac{s\left( \kappa \right) }{R} P_{\text{T}},
\end{equation}
where $s$ represents the bit length of the compressed prompt $\hat{\textit{\textbf{x}}}$ with compression ratio $\kappa$, $P_{\text{T}}$ is the transmit power, $R$ is the transmission rate that can be expressed as
\begin{equation}
    R = W\log_2 \left(1 + \gamma \right) = W\log_2 \left(1 + \frac{P_{\text{T}}gd^{-\alpha}}{\sigma^2} \right).
\end{equation}
Here, $W$ is the bandwidth of the offloading link between the user and DCO, $\gamma$ is the Signal-to-Noise-Ratio (SNR), $P_T$ denotes the transmission power, $g$ is the Rayleigh fading coefficient (exponentially distributed with unit mean), $d$ represents the distance between user and DCO, $\alpha$ is the path-loss exponent, $\sigma^2$ represents the Gaussian noise term in the Additive White Gaussian Noise (AWGN) channel.

\subsection{Service Delay}
The total time consumption $T$ comprises three components: encoding delay in SLM and LLM, and transmission delay, expressed as:
\begin{equation}
    T \left( \kappa, P_T \right) = t_{\text{e}}^{{\rm SLM}} \left( \kappa \right) + t_{\text{e}}^{{\rm LLM}} \left( \kappa \right) + t_{\text{t}}\left( \kappa, P_T \right).
\end{equation}

\section{Joint Power and Prompt Optimizatiom}
This section introduces the JPPO for wireless network-aided LLM services. After formulating the problem, we propose a Double Deep Q-Network (DQN) method~\cite{van2016deep} to address the joint optimization problem. 

\subsection{Problem Formulation}
For our JPPO framework, we formulate an optimization problem that balances three key aspects: prompt compression quality, wireless transmission efficiency, and LLM service performance. The objective is to maximize the overall fidelity while satisfying power and latency constraints in the wireless network-aided LLM service system. Specifically, the joint optimization problem can be formulated as
\begin{align}
\mathop
{\max}\limits_{\{ \kappa, P_T \}} &\
 \mathbf{f}\left(\kappa, \eta\left(P_T\right)\right), \label{equ:consuser}\\
\text{s.t.}
&~E \left(\kappa, P_T\right) \le E_{\text{th}}, \tag{\ref{equ:consuser}{a}}\label{equ:consuser d}\\
&~P_{\text{T}} \le P_{\text{th}}, \tag{\ref{equ:consuser}{b}}\label{equ:consuser a}\\
&~T \le T_{\text{th}},\tag{\ref{equ:consuser}{c}}\label{equ:consuser b}\\
&~\mathbf{f} > \mathbf{f}_{\text{th}},\tag{\ref{equ:consuser}{d}}\label{equ:consuser c}
\end{align}
where $\eta$ is the BER that is affected by the wireless environment and the transmit power, $P_{\text{th}}$ is the maximum allowable power consumption, $T_{\text{th}}$ represents the maximum tolerable end-to-end latency, $\mathbf{f}_{\text{th}}$ defines the minimum required fidelity. The constraints are designed to ensure practical system operation. 
Constraint~\eqref{equ:consuser d} is the energy constraint that represents the total energy budget limitation at the edge device side, introducing a critical trade-off: while higher transmission power $P_T$ can lead to lower BER $\eta$ and thus improved wireless transmission fidelity $\mathbf{f_2}$ to enhance the overall fidelity $\mathbf{f}$, the energy constraint forces a higher compression ratio $\kappa$ to reduce both the SLM/LLM inference energy cost and wireless transmission energy consumption; however, an excessively high compression ratio can result in significant information loss from the original prompt, potentially degrading both the semantic preservation fidelity $\mathbf{f_1}$ and LLM service quality fidelity $\mathbf{f_3}$.
Constraint~\eqref{equ:consuser a} ensures the power consumption remains within the device's power budget, Constraint~\eqref{equ:consuser b} guarantees that the total service latency meets real-time requirements, Constraint~\eqref{equ:consuser c} maintains the quality of service by enforcing a minimum threshold on fidelity. This optimization framework allows us to find the optimal balance between compression ratio and transmission power while maintaining high-quality LLM service delivery.

 
\subsection{Double DQN Solution}
\addtolength{\topmargin}{0.4in}
To solve the complex JPPO optimization problem, we deploy a centralized Double DQN method to find optimal prompt compression and transmission strategies for $N$ users, as shown in~\textbf{Algorithm 1}. The key elements of the proposed Double DQN design are
\begin{itemize}
\item \emph{Environment:} The environment of the Double DQN algorithm in the proposed framework is the communication environment with $N$ users.
\item \emph{State:} The state information includes the current fidelity of the transmitted message, SNR, and BER. The state information of $n_{\rm th}$ user is captured in a 3-dimensional vector: $\left[\mathbf{f}_n(\eta_n), \gamma_n, \right]$.
\item \emph{Action:} The actions include selecting compression and power levels. The action space is denoted as $\mathcal{A}=\{\mathcal{A}_1,\cdots, \mathcal{A}_N\}$, where $\mathcal{A}_n$ is the action of user $n$ and consists of a tuple with discrete values of compression ratio and transmission power level. The compression ratio level is discrete values range from $0$ to $4$. The transmission power level is discrete values ranging from $0$ to $9$, which affects BER.

\item \emph{Reward:} The reward of user $n$ is $\mathcal{R}_n$, and in each episode, the agent accumulates rewards based on the actions. The reward function maximizes fidelity while minimizing penalties related to BER and power usage.
\end{itemize}

The key design of the Double DQN is to decouple action selection from evaluation and address the over-estimation issue by using two separate networks:{\textit{Current Q-network}}, which predicts Q-values based on the current state, and {\textit{Target Q-network}}, which calculates target Q-values during updates and evaluates the Q-value of the best next action selected by the current Q-network, making the learning process of the Double DQN more stable.

\begin{algorithm}
\renewcommand{\algorithmicrequire}{\textbf{Input:}}

\renewcommand{\algorithmicensure}{\textbf{Output:}}
\caption{Double DQN Algorithm}
\label{table}
\begin{algorithmic}[1]
\REQUIRE  Initialize the  action space $\mathcal{A}=\{\mathcal{A}_1,\cdots, \mathcal{A}_N\}$ and target privacy $(\epsilon, \delta)$.

\ENSURE Reward.
\STATE \emph{Initialization:} $s_0$.
\FOR {each episode $k \in \{1,...,K\}$}
\STATE Explore actions and obtain initial states.
\FOR {each step in the episode}
\STATE Take action $a_t$ according to current policy.
\STATE Receive reward $r_t$ and observe the next state $s_{t+1}$.
\IF {$s'$ is the final state}
\STATE $y=\mathcal{R}'$.
\ELSE 
\STATE $y=\mathcal{R}'+[ \alpha r_{t+1}+ \gamma \max_{a} Q(s_{t+1}, a) - Q(s_t, a_t) ]$.
\ENDIF
\STATE Sample a mini-batch of experiences from memory
\FOR {each experience $(s, a, r, s')$ in the batch:}
\STATE Calculate target Q-values according to (13)
\STATE Compute the loss according to (12)
\STATE Perform gradient descent to minimize loss:
\STATE Update $\epsilon$: $\epsilon = \max\left(\epsilon * \epsilon_{\text{decay}}, \epsilon_{\min}\right)$
\STATE Update the target Q-network every few episodes.
\STATE Accumulate total reward and update state:
\STATE $\text{total reward} += r$, $s = s'$
\ENDFOR
\ENDFOR
\ENDFOR
\STATE Update the policy parameter.
\STATE Terminate the training when the policy converges or after a predefined number of iterations.
\end{algorithmic}
\end{algorithm}

The update steps of DQN are:  
\begin{equation} 
\begin{split}
&Q(s_t, a_t) \leftarrow Q(s_t, a_t) + [ \alpha r_{t+1}\\
&+ \mu \max_{a} Q(s_{t+1}, a) - Q(s_t, a_t) ], \end{split}  
\end{equation}  
where $ Q(s_t, a_t) $ is the estimated Q-value updated by the Bellman equation for taking action $ a_t $ in state $ s_t $. $ \alpha $ is the learning rate. $ r_{t+1} $ is the reward received after taking action $ a_t $ in state $ s_t $ and transitioning to state $ s_{t+1} $. $ \mu $ is the discount factor. $ \max_{a} Q(s_{t+1}, a) $ is the maximum estimated Q-value for all possible actions in state $ s_{t+1} $.  
    
In DQN, the loss function is the mean squared error (MSE) between the predicted Q-values and the target Q-values:  
\begin{equation} 
L_i(\theta_i) = \left[ r_{t+1} + \mu \max_{a'} Q_{\text{target}}(s_{t+1}, a'; \theta_i^*) - Q(s_t, a_t; \theta_i) \right]^2 , 
\end{equation}  
where $ L_i(\theta_i) $ is the loss for the $ i $-th iteration with parameter $ \theta_i $. $ r_{t+1}$ and $ s_{t+1}$ are the reward and next state observed after taking action $ a_t $ in state $ s_t $. $ Q_{\text{target}}(s_{t+1}, a'; \theta_i^*) $ is the target Q-value estimated by the target network with parameters $ \theta_i^* $.  $ Q(s_t, a_t; \theta_i) $ is the predicted Q-value by the current Q-network with parameters $ \theta_i $. 
  

The update steps of the Double DQN can be expressed as
\begin{equation}
    y = r + \mu \cdot Q_{\text{target}} \left( s', \arg\max_{a'} Q(s', a'; \theta); \theta^- \right),
\end{equation}
where $Q(s, a; \theta)$ is the estimated Q-value from the current Q-network with parameter $\theta$ and $Q_{\text{target}} ( s', a'; \theta^-)$ is the Q-value from the target network with parameter $\theta^-$.
\section{Numerical Results}
We design a customized environment with variable fidelity, SNR, and BER to simulate the wireless network-aided LLM service framework. 
The centralized DQN agent manages the environment, which involves selecting compression and power levels. Its goal is to balance fidelity, minimize errors, and optimize power usage. 
We employ the LLMLingua platform, utilizing the SLM model with GPT-Neo 125M to do prompt compression and use GPT-J 6B to generate the response~\cite{rothman2022transformers}. The simulations are carried out on MeetingBank-transcript dataset~\cite{hu2023meetingbankbenchmarkdatasetmeeting}. The parameter settings are listed in Table. 1.
\begin{table}
    \small
    \setlength{\tabcolsep}{1pt}
\caption{Simulation parameter configuration}
\begin{tabularx}{8cm}{lX}
\toprule

  {Parameter}  &  {Value}\\
 \midrule
  Learning rate & $10^{-3}$ \\
  $\alpha_1$, $\alpha_2$, and $\alpha_3$ & $0.4$, $0.3$ and $0.3$\\
  Total test runs range &$[1,10]$\\
  Episodes per test run & $10,000$\\

  \bottomrule
\end{tabularx}
\end{table}




\begin{figure*}[t!]
\centering
\includegraphics[width=0.83\textwidth]{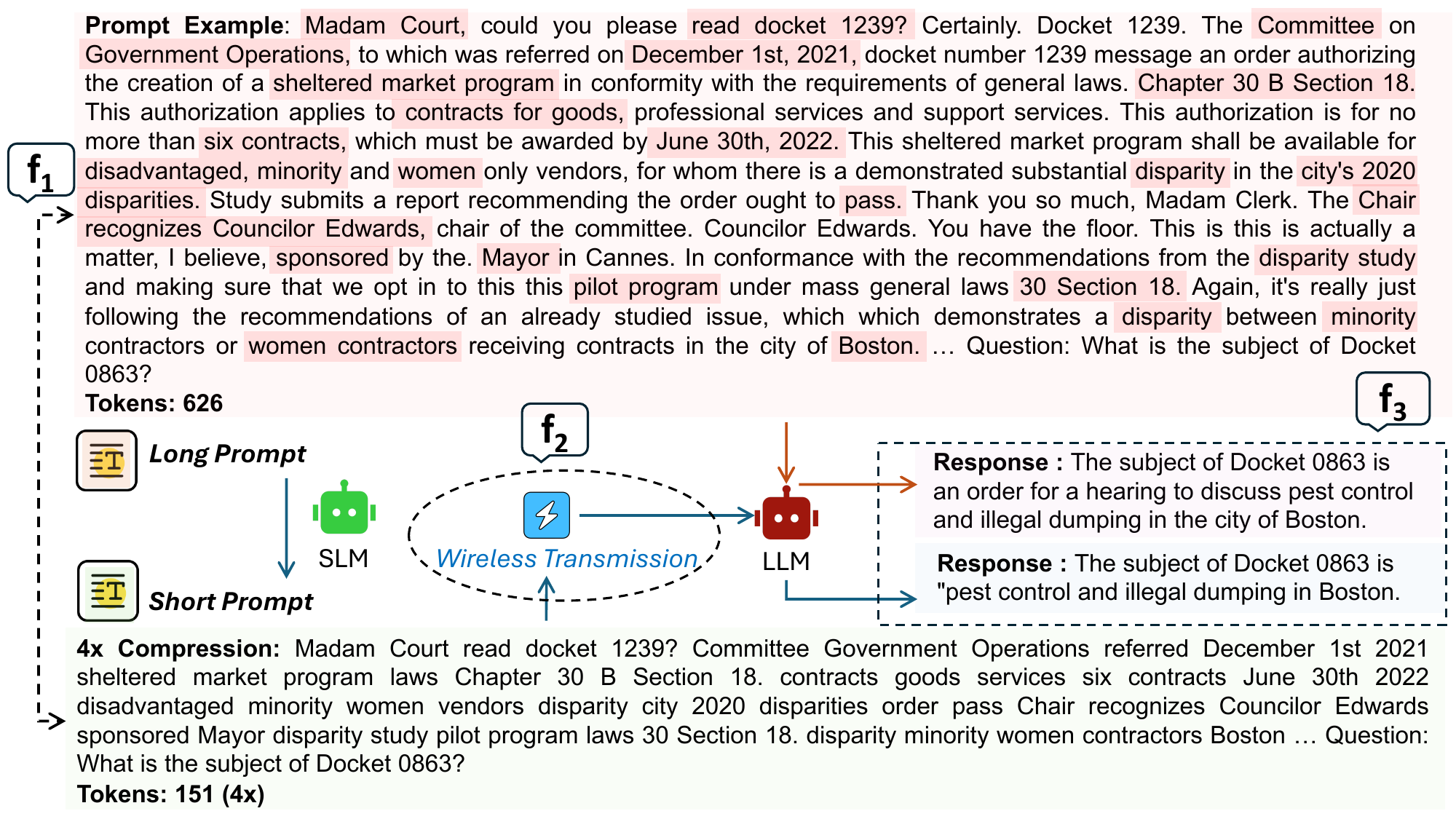}
   \caption{The example illustrates wireless network-aided LLM services with SLM-based prompt compression, with a 4x compression ratio. The highlighted parts represent key information from the original prompt. Additionally, we show the corresponding steps of measuring the three sub-performance metrics ($\mathbf{f_1}$, $\mathbf{f_2}$ and $\mathbf{f_3}$) of the fidelity metric $\mathbf{f}$ throughout the process.}
    \label{fig:update_example}
\end{figure*}


Fig.~\ref{fig:update_example} shows one example of prompt compression with a 4x compression ratio. We can observe that the length of the original prompt's text has been significantly reduced by the SLM. Furthermore, even with the significant reduction in prompt length, the accuracy of the responses from the LLM has not been compromised, indicating the effectiveness of the proposed SLM-based design.

\begin{figure}
\centering
\includegraphics[width=0.45\textwidth]{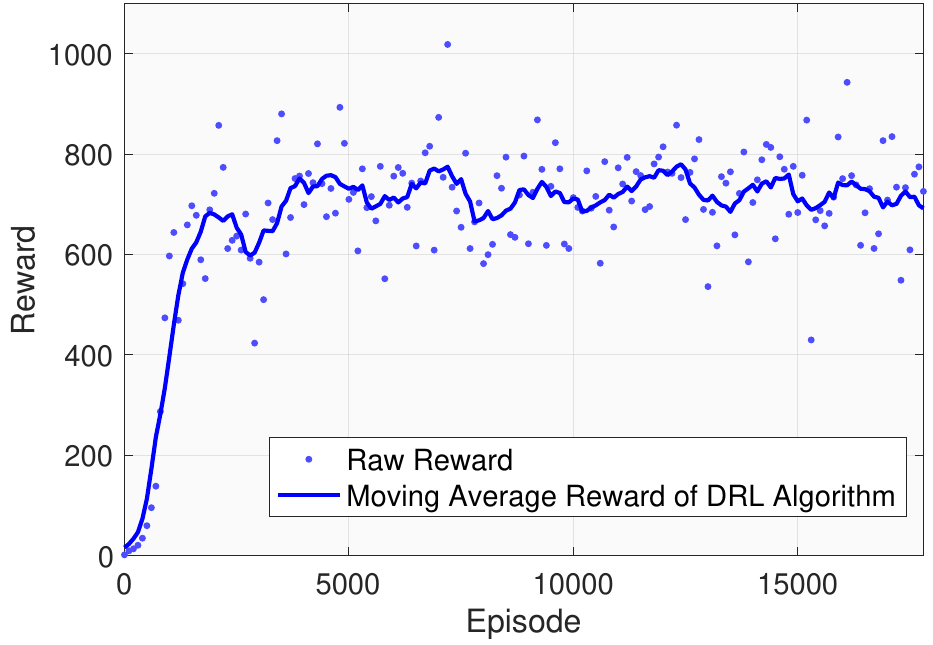}
  \caption{The convergence performance of reward for the proposed Double DQN algorithm.}
    \label{fig:reward}
\end{figure}
Fig. 4 shows the number of episodes and the reward convergence of the proposed Double DQN method when running $10,000$ episodes. We can observe from the figure that as training episodes progress, the reward increases over episodes with exploration and then stabilizes within the range of $700$ to $800$.
Using the optimal policy from our trained DRL model, we conducted end-to-end inference experiments. Specifically, for a short original prompt of $44$ tokens, the total response time was reduced from $56.1$ seconds to $46.9$ seconds. When testing with a longer prompt of $388$ tokens, the total response time decreased from $85.3$ seconds to $71.2$ seconds.
Further evaluation across our test dataset showed that the overall performance improvement fluctuates around $17\%$.

\begin{figure}[ht]
\centering
\includegraphics[width=0.45\textwidth]{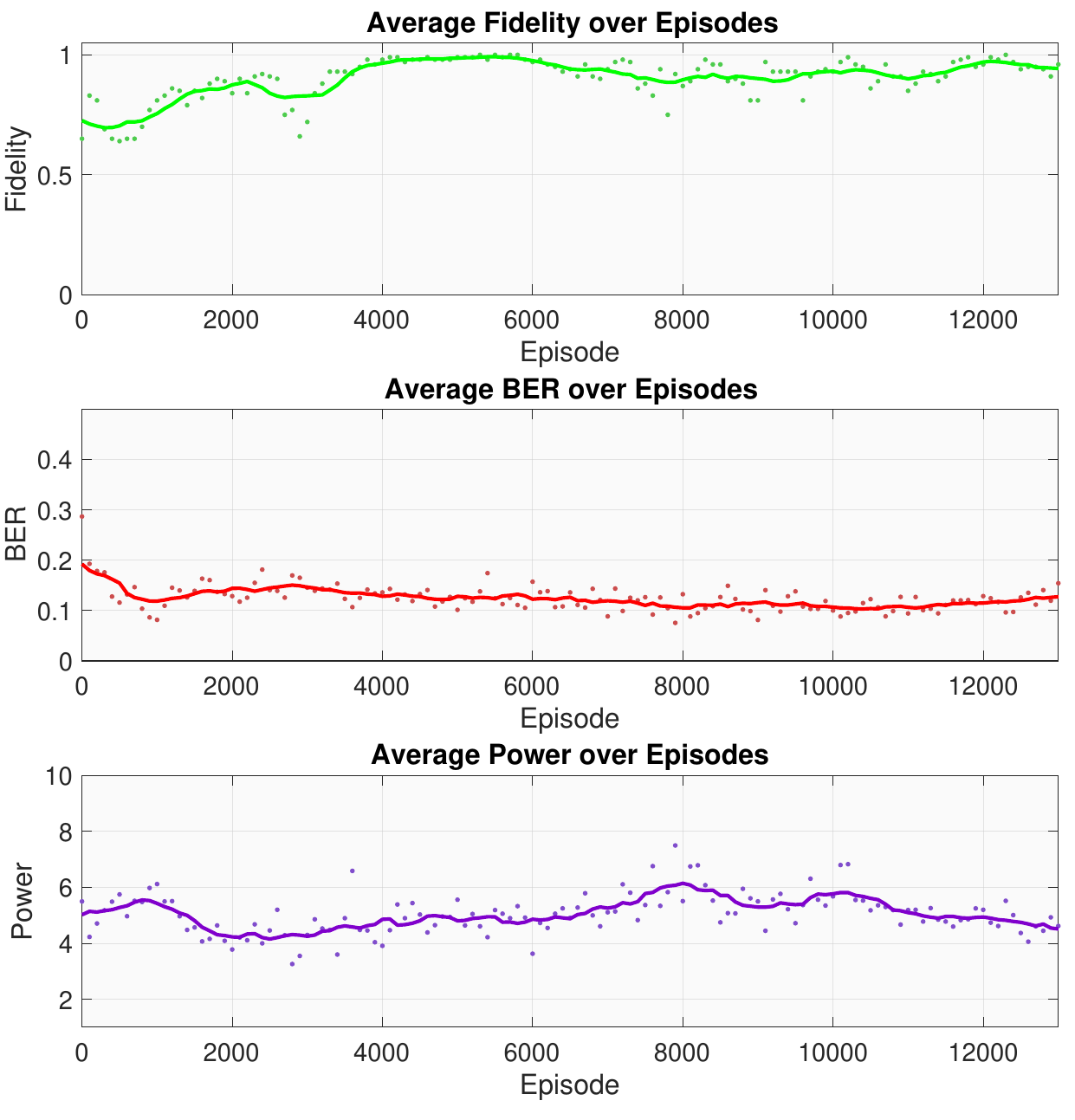}
   \caption{The training performance of the proposed Double DQN algorithm. Figs. 5(a), 5(b), and 5(c) show the fidelity performance, the BER, and the power consumption, respectively.
}
    \label{fig:reward}
\end{figure}
Fig. 5 shows the proposed algorithm's performance over training episodes.  
Fig. 5(a) plots the average fidelity over episodes. Fidelity measures the accuracy between the model's output and the reference before and after compression. The closer the fidelity value is to $1$, the more accurate it is after prompt compression, indicating that the compressed prompt retains most of the original information. 
We can observe that the fidelity increases with episode growth and maintains a range around $0.9$. This indicates that the proposed design is achieving high levels of accuracy after prompt compression. 
Fig. 5(b) plots the average BER over episodes and shows a decreasing trend across the episodes. The closer the BER is to $0$, the higher the accuracy that is achieved for the communication system. Even in our simulated harsh environment without error correction coding, It is shown that BER drops to a low level as episodes increase and remains below 0.2, demonstrating the system's effectiveness in minimizing transmission errors.
Fig. 5(c) plots the average power consumption over episodes. We can see that the power usage is well-regulated, staying within a range of $4$ ${\rm W}$ to $5$ ${\rm W}$. This controlled power consumption is particularly noteworthy given the inherent trade-offs: higher compression ratios can allow more transmit power for reliable transmission, while lower power may compromise transmission quality. The results in Fig. 5 demonstrate that our proposed algorithm effectively balances multiple objectives - maintaining high fidelity and low BER while keeping energy consumption within acceptable limits.

\section{Conclusion}
We proposed a novel power and prompt optimization framework for wireless network-aided accelerated LLM services. The proposed mechanism achieves effective performance in high service fidelity and relatively low BER, and the power consumption is maintained in constraint. Experimental evaluation shows that the proposed algorithm achieves stable convergence, indicating its potential for future practical deployment in LLM service systems.

\bibliographystyle{IEEEtran}
\bibliography{Ref}
\end{document}